# ANALYSIS OF RSA ALGORITHM USING GPU PROGRAMMING

Sonam Mahajan[1] and Maninder Singh[2]

[1]Department of Computer Science Engineering, Thapar University, Patiala, India
sonam_mahajan1990@yahoo.in

[2] Department of Computer Science Engineering, Thapar University, Patiala, India
msingh@thapar.edu

*ABSTRACT*

*Modern-day computer security relies heavily on cryptography as a means to protect the data that we have become increasingly reliant on. The main research in computer security domain is how to enhance the speed of RSA algorithm. The computing capability of Graphic Processing Unit as a co-processor of the CPU can leverage massive-parallelism. This paper presents a novel algorithm for calculating modulo value that can process large power of numbers which otherwise are not supported by built-in data types. First the traditional algorithm is studied. Secondly, the parallelized RSA algorithm is designed using CUDA framework. Thirdly, the designed algorithm is realized for small prime numbers and large prime number . As a result the main fundamental problem of RSA algorithm such as speed and use of poor or small prime numbers that has led to significant security holes, despite the RSA algorithm's mathematical soundness can be alleviated by this algorithm.*

## 1. INTRODUCTION

RSA (named for its inventors, Ron Rivest, Adi Shamir, and Leonard Adleman [1] ) is a public key encryption scheme. This algorithm relies on the difficulty of factoring large numbers which has seriously affected its performance and so restricts its use in wider applications. Therefore, the rapid realization and parallelism of RSA encryption algorithm has been a prevalent research focus. With the advent of CUDA technology, it is now possible to perform general-purpose computation on GPU [2]. The primary goal of our work is to speed up the most computationally intensive part of their process by implementing the GCD comparisons of RSA keys using NVIDIA's CUDA platform.

The reminder of this paper is organized as follows. In section 2,3,4, we study the traditional RSA algorithm. In section 5, we explained our system hardware. In section 6,7,8, 9 we explained the design and implementation of parallelized algorithm. Section 10 gives the result of our parallelized algorithm and section 12 concludes the paper.

## 2. TRADITIONAL RSA ALGORITHM[1]

RSA is an algorithm for public-key cryptography [1] and is considered as one of the great advances in the field of public key cryptography. It is suitable for both signing and encryption. Electronic commerce protocols mostly rely on RSA for security. Sufficiently long keys and up-to-date implementation of RSA is considered more secure to use.

RSA is an asymmetric key encryption scheme which makes use of two different keys for encryption and decryption. The public key that is known to everyone is used for encryption. The messages encrypted using the public key can only be decrypted by using private key. The key generation process of RSA algorithm is as follows:

The public key is comprised of a modulus n of specified length (the product of primes p and q), and an exponent e. The length of n is given in terms of bits, thus the term "8-bit RSA key" refers to the number of bits which make up this value. The associated private key uses the same n, and another value d such that d*e = 1 mod φ (n) where φ (n) = (p‐1)*(q‐1) [3]. For a plaintext M and cipher text C, the encryption and decryption is done as follows:

$$C = M^e \bmod n, \quad M = C^d \bmod n.$$

For example, the public key (e, n) is (131,17947), the private key (d, n) is (137,17947), and let suppose the plaintext M to be sent is: *parallel encryption*.
- Firstly, the sender will partition the plaintext into packets as: pa ra ll el en cr yp ti on. We suppose a is 00, b is 01, c is 02, ..... z is 25.
- Then further digitalize the plaintext packets as: 1500 1700 1111 0411 0413 0217 2415 1908 1413.
- After that using the encryption and decryption transformation given above calculate the cipher text and the plaintext in digitalized form.
- Convert the plaintext into alphabets, which is the original: *parallel encryption*.

## 3. OPERATION [1]

The three steps of RSA algorithm – Key Generation, Encryption and Decryption are explained as follows:

### 3.1 Key generation

As explained above RSA is an asymmetric key encryption scheme. It makes use of two different keys for encryption and decryption. The public key that is known to everyone is used for encryption. The messages encrypted using the public key can only be decrypted by using private key. The keys for RSA algorithm are generated as follows:

**RSA KEY GENERATION**
- Choose two different random prime numbers say p and q. From security point of view both integers should be chosen at random and of similar size(bit length);
- Compute the value of n = p*q. The length of n is the actual key length usually expressed in bits and n is used as modulus for both public and private keys;
- Compute $\varphi$ (Euler's totient function). $\varphi(n) = \varphi(p)\varphi(q) = (p-1)(q-1)$;
- Choose an integer e such that e lies between 1 and $\varphi(n)$ (1<e< $\varphi(n)$). Make sure e and $\varphi(n)$ are co prime i.e., gcd(e, $\varphi(n)$) = 1 ;
- Calculate the value of d such that $d*e = 1(\mod \varphi(n))$;
- Publish public key as (e, $\varphi(n)$). It should be publically available.
- Keep private key as (d, $\varphi(n)$) secret along with the values of p, q, $\varphi(n)$.

Figure 1. RSA Key Generation

### 3.2 Encryption process

Alice transmits its public key (n,e) to the Bob and keeps the private key secret. Suppose Bob wishes to send message M to Alice. He first convert the message M to an integer m , such that m is between 0 and n using padding scheme [1][3]. Finally cipher text c is calculated as: $C = m^e \mod n$. This cipher text is then transmitted to Alice.

### 3.3 Decryption process

Alice can decode the cipher text using her private key component d as follows:

$m = C^d \mod n$

Given m, original message M can be recovered.

## 3.4 Example

**RSA EXAMPLE**

- Select primes : p =17 and q =11;
- Compute n=p q = 17*11 =187;
- Compute φ(n) = (p-1)(q-1) = 16*10=160;
- Select e: gcd(e,160) =1; choose e = 7;
- Compute d : de = 1 mod 160 and d <160 , d=23;
- Publish public key {7,187};
- Keep private key secret {23,17,11}

Figure 2. RSA Example

## 3.6 Modular exponential

### 3.6.1 Modular arithmetic

Public key cryptography is computationally expensive as it mostly includes raising a large power to a base and then reducing the result using modulo function. This process is known as modular exponential. In practical, RSA algorithm should be fast , i.e., modular exponential function should be fast and along with it should be efficient. The simple modular operations are given in the Figure 3.

$$(u + v) \bmod m = ((u \bmod m) + (v \bmod m)) \bmod m$$
$$(u - v) \bmod m = ((u \bmod m) - (v \bmod m)) \bmod m$$
$$(u * v) \bmod m = ((u \bmod m) * (v \bmod m)) \bmod m$$

Figure 3. Modular Arithmetic

### 3.6.2 Naive modular exponentiation

In this method modular multiplications are applied repeatedly. For example: with g=4, e=13 and m= 497, the naive modular exponential will solve ge mod m as shown in figure and gives the result as 445. This method is not efficient as it performs e-1 modular multiplications. In cryptography the security depends on the larger value of e and efficiency depends how efficiently these modular multiplications and modular exponential functions are solved. The plaintext , cipher-text or even partial cipher text is supposed to be of large in value and it will require a large amount of modular multiplications if we rely on this naive algorithm.

1. e=1, c= 4 mod 497 = 4.
2. e=2, c=(4 _ 4) mod 497 = 16 mod 497 = 16.
3. e=3, c=(16 _ 4) mod 497 = 64 mod 497 = 64.
4. e=4, c=(64 _ 4) mod 497 = 256 mod 497 = 256.
5. e=5, c=(256 _ 4) mod 497 = 1024 mod 497 = 30.
6. e=6, c=(30 _ 4) mod 497 = 120 mod 497 = 120.
7. e=7, c=(120 _ 4) mod 497 = 480 mod 497 = 480.
8. e=8, c=(480 _ 4) mod 497 = 1920 mod 497 = 429.
9. e=9, c=(429 _ 4) mod 497 = 1716 mod 497 = 225.
10. e=10, c=(225 _ 4) mod 497 = 900 mod 497 = 403.
11. e=11, c=(403 _ 4) mod 497 = 1612 mod 497 = 121.
12. e=12, c=(121 _ 4) mod 497 = 484 mod 497 = 484.
13. e=13, c=(484 _ 4) mod 497 = 1936 mod 497 = 445.

Figure 4. Naive Modular Exponentiation Example

### 3.6.3 Repeated square-and-multiply methods

It makes use of the fact that if the e value is even, then the modular exponential is calculated as ge mod m =( ge/2 * ge/2) mod m ,hence by reducing the amount of modular multiplications to 2z where z is the number of bits when converting e to binary form. The algorithm comes with two forms as shown below.

- Right-to-left binary modular exponential
- Left-to-right binary modular exponential

This algorithm works efficiently for large value of e.

**Algorithm: Right-to-left binary modular exponentiation**

Input: an element g and integer $e \geq 1$, and a modulus m.
Output: $g^e$ mod m.

1. $A = 1$; $S = g$; $E = e$.
2. While $E \neq 0$ do the following:
   2.1. If E is odd, then $A = (A \cdot S)$ mod m, $E = E - 1$.
   2.2. $E = E/2$.
   2.3. If $E \neq 0$, then $S = (S \cdot S)$ mod m.
3. Return ( A ).

**Algorithm: Left-to-right binary modular exponentiation**

Input: an element g and a positive integer $e = (e_t\ e_{t-1}\ \_\_\_\ e_1\ e_0)_2$ ,and a modulus m.
Output: $g^e$ mod m.

1. $A = 1$.
2. For i from t down to 0 do the following:
   2.1. $A = (A \cdot A)$ mod m.
   2.2. If $e_i = 1$, then $A = (A \cdot g)$ mod m.
3. Return ( A ).

Figure 5. Repeated Square-and-Multiply Methods

### 3.6.4 Left-to-right k-ary exponentiation

This algorithm is the generalization of the previous left-to-right binary modular exponentiation described previously. In this algorithm each iteration processes more than one bit of the exponent and works efficiently when pre-computations are done in advance and is used again and again.

```
Algorithm: Left-to-right k-ary modular exponentiation

Input: g and e =(e_t e_{t-1} _ _ _ e_1 e_0)_b, where b=2^k for some k≥1 and a modulus m.
Output: g^e mod m

1. Precomputation.
1.1. g_0 = 1.
1.2. For i from 1 to (2^k -1)  do: g_i =(g_{i-1} .g)

2. A = 1.

3. For i from t down to 0 do the following:
3.1. A = ( A^{2^k}) mod m
3.2. A = ( A .g_{e_i}) mod m.

4. Return ( A ).
```

Figure 6. Left-To-Right K-ary Modular Exponentiation

### 3.6.5 Sliding window exponential

The main idea behind this algorithm is the reduction of pre-computations as compared to the above discussed algorithm and hence ultimately reduction in the average number of multiplications done.

```
Algorithm: Sliding-window exponential

Input: g and e = ( e_t e_{t-1}..........e_1 e_0)_2 , with e_t = 1, an integer k≥ 1, and a modulus m.

Output: g^e mod m.

1. Precomputation.

    1.1. g_1 = g, g_2 = g^2

    1.2. For i from 1 to (2^{k-1} – 1) do: g_{2i+1} = ( g_{2i – 1} . g_2) mod m.

2. A = 1, i = t

3. While i ≥ 0 do the following:

    3.1. If e_i = 0 then do: A = A^2 mod m , i = i-1.

    3.2. Otherwise (e_i ≠ 0) , find the longest bitstring e_i e_{i-1}.......e_l such that i-l+1 ≤ k and e_l =1, and do the following: A=A^{2^{i-l+1}} . g_{(e_i e_{i-1}.......e_l)_2} , i=l-1.

4. Return (A).
```

Figure 7. Sliding Window Exponential

### 3.7 ADVANTAGES AND DISADVANTAGES OF RSA ALGORITHM

The RSA algorithm is known for its increased security and convenience. In RSA algorithm private keys are neither transmitted or nor revealed to anyone. By contrast in secret-key system keys are exchanged either manually or with the help of communication channel. This is because in secret-key system same key is used for encryption and decryption and there exist a chance that an enemy can retrieve this secret key during transmission

Public-key systems can provide digital signatures. These digital signatures cannot be repudiated. By contrast in secret-key systems authentication requires either sharing of some important secrets or the involvement of third party. This can result in sender repudiating the previously

authenticated message by claiming that the shared secret is compromised any one of the party involved

Speed is the main disadvantage of public-key cryptography. Because there is always a trade off between efficiency and security. For efficient RSA, key size should be small but small key size leads to many security holes. There exist many secret-key methods that are faster than the asymmetric encryption method. In reality public-key encryption is used to supplement the private-key encryption.

## 4. GPU-ACCELERATED COMPUTING[4][3]

Cryptographic operations are highly computationally expensive and hence consume a large amount of CPU cycles from the system. Such examples are web servers. To incorporate with this and to offload cryptographic operations, web servers are typically deployed in systems using cryptographic accelerator cards. Though this saves system CPU cycles from complex and computationally intensive application logic but it also results in difficult deployment scenarios. So, the possibility of offloading the massive computationally rich operations to GPU would lighten the CPU load to a great extent and hence the saving can be used for other applications. The idea also provide a significant performance increase. This technique is known as "GPU ACCELERATED COMPUTING". The idea will be clear from the Figure. 8

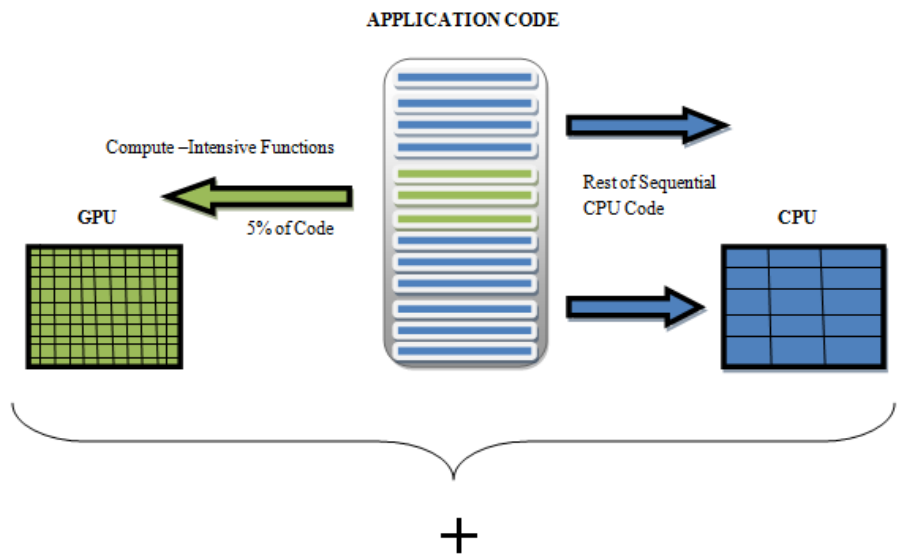

Figure 8. GPU Accelerated Computing

## 5. ARCHITECTURE OVERVIEW[4]

NVIDIA's Compute Unified Device Architecture (CUDA) platform provides a set of tools to write programs that make use of NVIDIA's GPUs [3]. These massively-parallel hardware devices process large amounts of data simultaneously and allow significant speedups in programs with sections of parallelizable code making use of the Simultaneous Program, Multiple Data (SPMD) model.

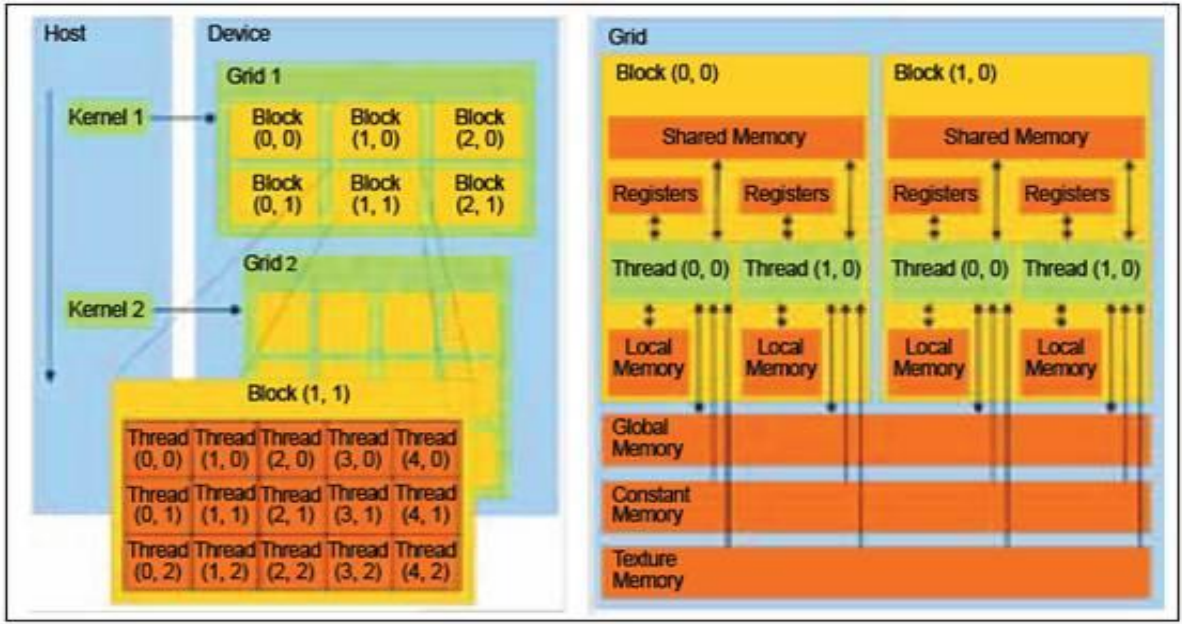

Figure 9. CUDA System Model

The platform allows various arrangements of threads to perform work, according to the developer's problem decomposition. In general, individual threads are grouped into up-to 3-dimensional blocks to allow sharing of common memory between threads. These blocks can then further be organized into a 2-dimensional grid. The GPU breaks the total number of threads into groups called warps, which, on current GPU hardware, consist of 32 threads that will be executed simultaneously on a single streaming multiprocessor (SM). The GPU consists of several SMs which are each capable of executing a warp. Blocks are scheduled to SMs until all allocated threads have been executed. There is also a memory hierarchy on the GPU. Three of the various types of memory are relevant to this work: global memory is the slowest and largest; shared memory is much faster, but also significantly smaller; and a limited number of registers that each SM has access to. Each thread in a block can access the same section of shared memory.

## 6. PARALLELIZATION

The algorithm used to parallelize the RSA modulo function works as follows:
- CPU accepts the values of the message and the key parameters.
- CPU allocates memory on the CUDA enabled device and copies the values on the device
- CPU invokes the CUDA kernel on the GPU
- GPU encrypts each message character with RSA algorithm with the number of threads equal to the message length.
- The control is transferred back to CPU
- CPU copies and displays the results from the GPU.

As per the flow given above the kernel is so built to calculate the cipher text $C = M^e \bmod n$. The kernel so designed works efficiently and uses the novel algorithm for calculating modulo value. The algorithm for calculating modulo value is implemented such that it can hold for very large power of numbers which are not supported by built in data types. The modulus value is calculated using the following principle:

- *$C = M^e \bmod n$*
- *$C = (M^{e-x} \bmod n * M^x \bmod n) \bmod n$*

Hence iterating over a suitable value of x gives the desired result.

# 7. ALGORITHM FOR CUDA KERNEL

**Algorithm 1:**

INPUT: Function to calculate modular exponential with *num,*key,*n and *results as input integer variables to __global__ void rsa()
OUTPUT : $num^{key}$ mod n

step 1: Start
step 2: Declare integer variables temp, total_no_of_threads
Step 3: Assign thread index
  i <- threadIdx.x + blockDim.x*blockIdx.x
Step 4: If ( i <- total_no_of_threads)
  temp <- mod(num[i],*key,*n)
  atomicExch(&result[i],temp)
Step 5:END

**Algorithm 2:**

INPUT: Function to calculate modular exponential with g,e,n as input integer variables to __device__ long long int mod()
OUTPUT: $g^e$ mod n

Step 1: Start
Step 2: Declare variables a,ret,size
Step 3: a <- (g%n) * (g%n)
Step 4: ret <- 1
Step 5: size <-e/2
Step 6: If (e==0)
    return (g%n)
  Else
    while(true)
      if(size>0.5)
      ret <-(ret*a)% n
      size <-size-1.0

      else if(size==0.5)
      ret <-(ret*(g%n))%n
      break

      else
      break

    return ret
Step 7:END

In the first algorithm we, first pass the arguments as pointer to global rsa(). After declaration of the variables we assign the thread index. Thread index is the unique id of the each thread in a block that will execute the same instruction but with different data sets that will be distributed among the threads. After that in the 4$^{th}$ Step, check the condition if the number of threads to be executed is greater than the threads assigned, call Algorithm 2 for mod() with three parameters. Finally after the execution of the Algorithm 2, encrypted text will be handled to the host as result[] array.

In Algorithm 1 __global__ keywords represents that the rsa() will execute on the device and will be called from the host. We use pointers for the variables because rsa() runs on device, so variables must point to device memory. For that we need to allocate memory on GPU

This algorithm works for large size of values which are not otherwise supported by the built-in data type.

## 8. Integration of C with CUDA API

```
cudaGetDeviceCount(&devcount);
printf("%d CUDA devices found",devcount);
if(devcount>0)
cudaSetDevice(1);
int *dev_num,*dev_key,*dev_den;
unsigned int *dev_res;
unsigned int res[3]={1,1,1};
cudaMalloc( (void **)&dev_num, nsize*sizeof(int));
cudaMalloc( (void **)&dev_key,sizeof(int));
cudaMalloc( (void **)&dev_den, sizeof(int));
cudaMalloc( (void **)&dev_res, nsize*sizeof(unsigned int));
ts = GetTickCount();
cudaMemcpy(dev_num,num,nsize*sizeof(int),cudaMemcpyHostToDevice);
cudaMemcpy(dev_key,&key,sizeof(int),cudaMemcpyHostToDevice);
cudaMemcpy(dev_den,&den,sizeof(int),cudaMemcpyHostToDevice);
cudaMemcpy(dev_res,res,nsize*sizeof(unsigned int),cudaMemcpyHostToDevice);
rsa<<<nblocks,nthreads>>>(dev_num,dev_key,dev_den,dev_res);
cudaMemcpy(res,dev_res,nsize*sizeof(unsigned int),cudaMemcpyDeviceToHost);
cudaFree(dev_num);
cudaFree(dev_key);
cudaFree(dev_den);
cudaFree(dev_res);
```

Figure 10. CUDA-C Host Code for RSA Algorithm

The above is the small part of the host code showing the integration of CPU and GPU.

- Check for available GPU devices;
- Initialize host and device copies of variables;
- Allocate space for device copies of variables using cudaMalloc();
- Setup input values;
- Copy input values to device using cudaMemcpy;
- Launch kernel on GPU;
  rsa<<<nblocks, nthreads>>>(dev_num,dev_key,dev_den,dev_res)
- Copy result back to host using cudaMemcpy();
- Cleanup.

Finally after the execution of the Algorithm 2, encrypted text will be handled to the host as result[] array.

In Algorithm 1 __global__ keywords represents that the rsa() will execute on the device and will be called from the host. We use pointers for the variables because rsa() runs on device, so variables must point to device memory. For that we need to allocate memory on GPU

This algorithm works for large size of values which are not otherwise supported by the built-in data type.

## 9. KERNEL CODE

As introduced in section 2, RSA algorithm divides the plaintext or cipher text into packets of same length and then apply encryption or decryption transformation on each packet. A question is how does a thread know which elements are assigned to it and are supposed to process them? CUDA user can get the thread and block index of the thread call it in the function running on device. In this level, the CUDA Multi-threaded programming model will dramatically enhanced the speed of RSA algorithm. The experimental results will be showed in section 10.

The kernel code used in our experiment is shown below. First CUDA user assign the thread and block index, so as to let each thread know which elements they are supposed to process. It is shown in Figure 11. Then it call for another device function to calculate the most intense part of the RSA algorithm. Note in the below figure11 and figure12, it works for 3 elements.

```
__global__ void rsa(int * num, int *key, int *den, unsigned int * result)
{
    int i=threadIdx.x;
    int temp;
    if(i<3)
    {
        temp=mod(num[i],*key,*den);
        atomicExch(&result[i],temp);
    }
}
```

Figure 11. Kernel code

```
__device__ long long int mod(int base, int exponent, int den)
{
    unsigned int a=(base%den)*(base%den);
    unsigned long long int ret=1;
    float size=(float)exponent/2;
    if(exponent==0)
    {
        return base%den;
    }
    else
    {
        while(1)
        {
            if(size>0.5)
            {
                ret=(ret*a)%den;
                size=size-1.0;
            }
            else if(size==0.5)
            {
                ret=(ret*(base%den))%den;
                break;
            }
            else
            {
                break;
            }
        }
        return ret;
    }
}
```

Figure 12. Kernel's Device code

# 10. VERIFICATION

In this section we setup the test environment and design three tests. At first test, we develop a program running in traditional mode for small prime numbers (only use CPU for computing). And at the second test, we use CUDA framework to run the RSA algorithm for small prime numbers in multiple-thread mode. Comparison is done between the two test cases and speed up is calculated. In the third test we run the GPU RSA for large prime numbers that is not supported by the built-in data types of CPU RSA. The test result will be showed in this section

## 10.1. Test environment

The code has been tested for :

- Values of message between 0 and 800 which can accommodate the complete ASCII table
- 8 bit Key Values

The computer we used for testing has an Intel(R) Core(TM) i3-2370M 2.4GHZ CPU, 4 GB RAM, Windows 7OS and a Nvidia GeForce GT 630M with 512MB memory, and a 2GHZ DDR2 memory. At the first stage, we use Visual Studio 2010 for developing and testing the traditional RSA algorithm using C language for small prime numbers. Series of input data used for testing and the result will be showed later.
At the second stage, we also use Visual Studio 2010 for developing and testing parallelized RSA developed using CUDA v5.5 for small prime numbers. After that the results of stage one and stage second are compared and hence calculating the respective speedup.

In the third test we run the GPU RSA for large prime numbers that is not supported by the built-in data types of CPU RSA. The test result will be showed in this section. At present the calculation of Cipher text using an 8-bit key has been implemented parallel on an array of integers.

## 10.2 Results

In this part, we show the experiment results for GPU RSA and CPU RSA for small value of n.

Table 1. Comparison of CPU RSA and GPU RSA for small prime numbers i.e (n=131*137)

| Data Size(bytes) | No. of blocks | Threads per block | GPU RSA Time | CPU RSA Time | Speedup |
|---|---|---|---|---|---|
| 256 | 4 | 64 | 7.56 | 12.56 | 1.66 |
| 512 | 8 | 64 | 7.25 | 19.14 | 2.65 |
| 1024 | 16 | 64 | 6.86 | 23.60 | 3.44 |
| 2048 | 32 | 64 | 5.38 | 29.33 | 5.51 |
| 4096 | 64 | 64 | 5.68 | 32.64 | 5.74 |
| 8192 | 128 | 64 | 6.27 | 35.16 | 5.60 |
| 16392 | 256 | 64 | 7.21 | 39.66 | 5.50 |
| 32784 | 512 | 64 | 9.25 | 42.37 | 4.58 |

Table 1 shows the relationship between the amount of data inputting to the RSA algorithm and the execution times (in seconds) in traditional mode and multiple thread mode. The first column shows the number of the data input to the algorithm, and the second column shows the number of blocks used to process the data input. In the above table 64 threads per block are used to execute RSA. The execution time is calculated in seconds. In the last column speed up is calculated. Above results are calculated by making average of the results so taken 20 times to have final more accurate and precise results.

The enhancement of the execution performance using CUDA framework can be visually demonstrated by Figure 13.

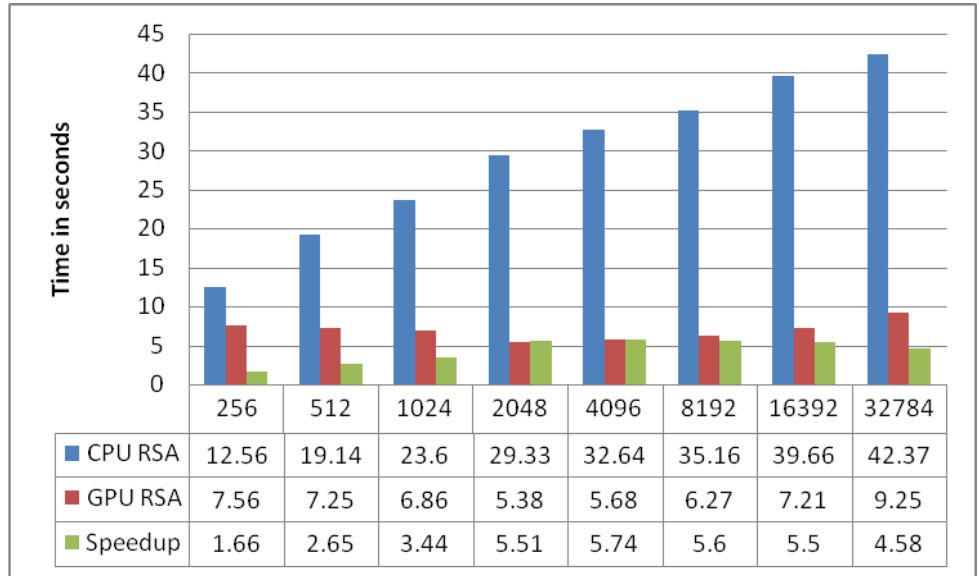

Figure 13. Graph showing effect of data input on CPU RSA and GPU RSA along with the Speedup

## 10.2.1 GPU RSA for large prime numbers

In this part, we show the experiment results for GPU RSA and CPU RSA for small value of n.

Table 2. GPU RSA for large prime numbers and large value of n (n = 1005 * 509)

| Data Size(bytes) | No. of blocks | Threads per block | GPU RSA Time |
|---|---|---|---|
| 256 | 8 | 32 | 6.08 |
| 512 | 16 | 32 | 6.52 |
| 1024 | 32 | 32 | 6.69 |
| 2048 | 64 | 32 | 5.53 |
| 4096 | 128 | 32 | 6.58 |
| 8192 | 256 | 32 | 6.66 |
| 16392 | 512 | 32 | 7.81 |
| 32784 | 1024 | 32 | 8.76 |

From Table 2, we can see the relationship between the execution time in seconds and the input data amount (data in bytes) is linear for certain amount of input. When we use 256 data size to execute the RSA algorithm, the execution time is very short as compared to traditional mode which is clearly proved in the above section where the comparison is made for CPU RSA and GPU RSA for small prime numbers and hence for small value of n. So we can say when the data size increases, the running time will be significantly reduced depending upon the number of threads used. Furthermore, we also find that when the data size increases from 1024 to 8192, the execution time of 7168 threads almost no increase, which just proves our point of view, the more the data size is, the higher the parallelism of the algorithm, and the shorter the time spent. Execution time varies according the number of threads and number of blocks used for data input. In the above table threads per block are constant i.e we use 32 threads per block and number of blocks used are adjusted according to the data input.

The enhancement of the execution performance of data input in bytes using the large value of prime numbers (n=1009 * 509) and hence large value of n on CUDA framework can be visually demonstrated by Figure 14.

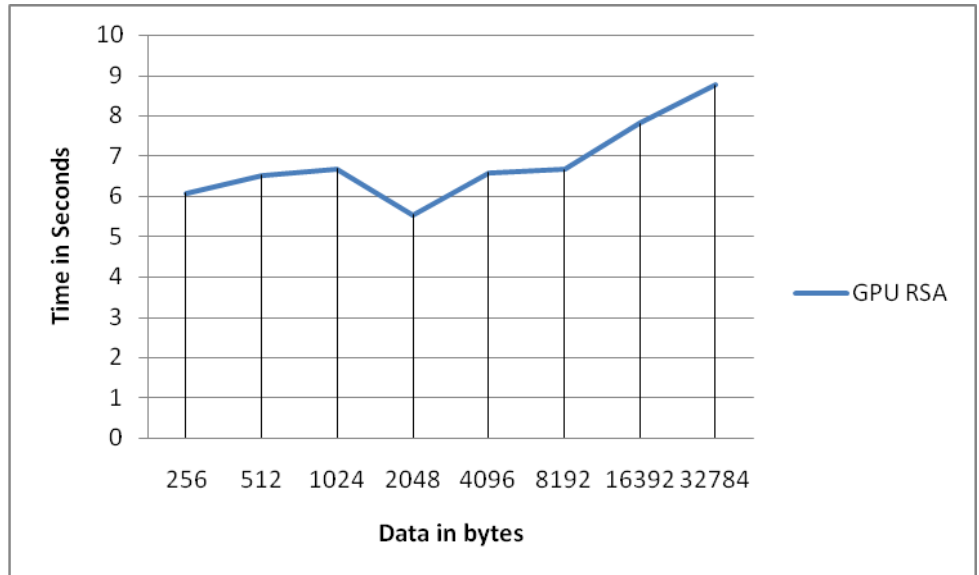

Figure 14. GPU RSA for large value of n (n=1009*509)

### 10.2.2. Execution time comparison of GPU RSA for large value of n (1009*509) with CPU RSA for small value of n(137*131)

In the third and final stage of test results analysis, we analyse our results between sequential RSA that is using small value of n (17947) and parallelized RSA that is making use of large prime numbers and large value of n (513581). The enhancement of the GPU execution performance of data input in bytes using the large value of prime numbers (n=1009 * 509) on CUDA framework and CPU RSA using small value of prime numbers (n=137*131) can be visually demonstrated by Figure 15. Hence, we can leverage massive-parallelism and the computational power that is granted by today's commodity hardware such as GPUs to make checks that would otherwise be impossible to perform, attainable.

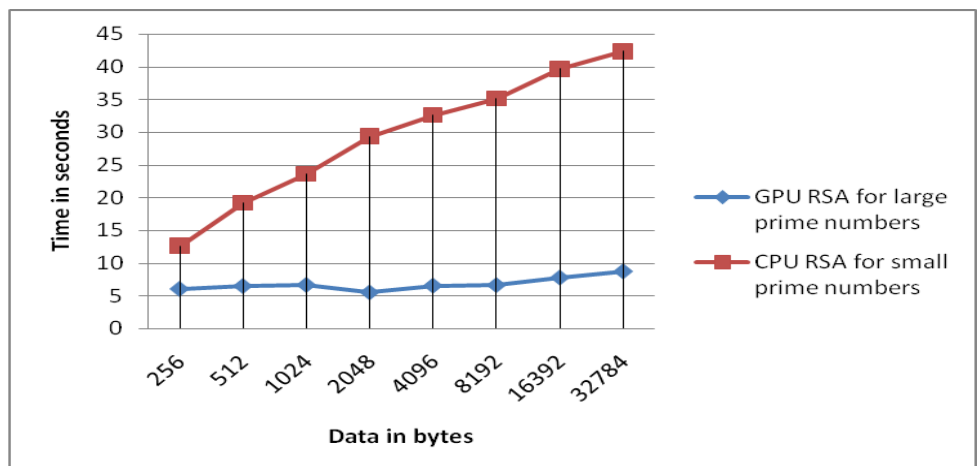

Figure 15. Comparison of CPU RSA for small prime numbers with GPU RSA for large prime numbers.

## 11. RSA DECRYPTION USING CUDA-C

In this paper, we presented our experience of porting RSA encryption algorithm on to CUDA architecture. We analyzed the parallel RSA encryption algorithm. As explained above the encryption and decryption process is done as follows:

$$C = M^e \mod n, \quad M = C^d \mod n.$$

The approach used for encryption process is same for decryption too. Same kernel code will work for decryption too. The only parameters that will change is the private key (d) and ciphertext in place of message bits used during encryption.

## 12. CONCLUSIONS

In this paper, we presented our experience of porting RSA algorithm on to CUDA architecture. We analyzed the parallel RSA algorithm. The bottleneck for RSA algorithm lies in the data size and key size i.e the use of large prime numbers. The use of small prime numbers make RSA vulnerable and the use of large prime numbers for calculating n makes it slower as computation expense increases. This paper design a method to computer the data bits parallel using the threads respectively based on CUDA. This is in order to realize performance improvements which lead to optimized results.

In the next work, we encourage ourselves to focus on implementation of GPU RSA for large key size including modular exponentiation algorithms. As it will drastically increase the security in the public-key cryptography. GPU are becoming popular so deploying cryptography on new platforms will be very useful.

**Authors**

Sonam Mahajan                                            Patiala-147004
Student , ME – Information Security
Computer Science and Engineering Department
Thapar University
Patiala-147004

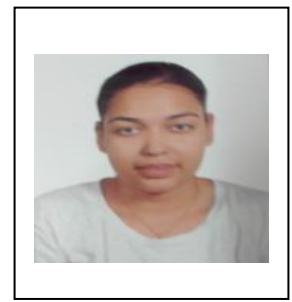

Dr. Maninder Singh
Associate Professor
Computer Science and Engineering Department
Thapar University

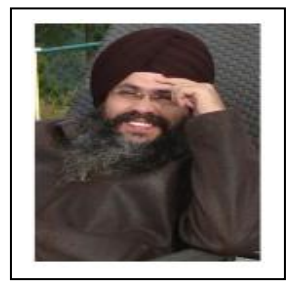